\documentclass[twocolumn, nobibnotes, aps, 10pt, superscriptaddress, prl, longbibliography]{revtex4-1}
\usepackage{graphicx}
\usepackage{amsmath}
\usepackage{multirow}
\usepackage{longtable}
\usepackage{epstopdf}
\usepackage{color}

\usepackage[colorlinks=true, allcolors=blue]{hyperref}
\usepackage{hyperref}

\bibpunct{[}{]}{,}{n}{}{}

\begin{document}

\title{Gate-tunable infrared plasmons in electron-doped single-layer antimony}

\author{D.~A. Prishchenko}
\email[]{d.a.prishchenko@urfu.ru}
\affiliation{\mbox{Theoretical Physics and Applied Mathematics Department,
Ural Federal University, Mira Str. 19, 620002 Ekaterinburg, Russia}}

\author{V.~G. Mazurenko}
\affiliation{\mbox{Theoretical Physics and Applied Mathematics Department,
Ural Federal University, Mira Str. 19, 620002 Ekaterinburg, Russia}}

\author{M.~I. Katsnelson}
\affiliation{\mbox{Theoretical Physics and Applied Mathematics Department,
Ural Federal University, Mira Str. 19, 620002 Ekaterinburg, Russia}}
\affiliation{\mbox{Institute for Molecules and Materials, Radboud University, Heijendaalseweg 135, 6525 AJ Nijmegen, The Netherlands}}

\author{A.~N. Rudenko}
\affiliation{School of Physics and Technology, Wuhan University, Wuhan 430072, China}
\affiliation{\mbox{Theoretical Physics and Applied Mathematics Department, Ural Federal University, 
Mira Str. 19, 620002 Ekaterinburg, Russia}}
\affiliation{\mbox{Institute for Molecules and Materials, Radboud University, Heijendaalseweg 135, 6525 AJ Nijmegen, The Netherlands}}

\date{\today}

\begin{abstract}
We report on a theoretical study of collective electronic excitations in single-layer antimony crystals (antimonene), a novel two-dimensional semiconductor with strong spin-orbit coupling. Based on a tight-binding model, we consider electron-doped antimonene and demonstrate that the combination of spin-orbit effects with external bias gives rise to peculiar plasmon excitations in the mid-infrared spectral range. These excitations are characterized by low losses and negative dispersion at frequencies effectively tunable by doping and bias voltage. The observed behavior is attributed to the spin-splitting of the conduction band, which induces interband resonances, affecting the collective excitations. Our findings open up the possibility to develop plasmonic and optoelectronic devices with high tunability, operating in a technologically relevant spectral range.
\end{abstract}


\maketitle

The growing field of plasmonics continues to gather attention from the material science community. Collective oscillations of electron density provide a way to couple indecent electromagnetic radiation to matter, which enables one to confine and enhance local field inside the material, essentially turning optical signal to electrical. Their practical use is diverse and depends on the desirable frequency region. The related fields include biosensing, light harvesting, optical thermal heating, lasers, photodetection and others~\cite{Maier,Murray,Boriskina,kneipp1997,mayer2011,rodrigo2015,sorger2012}. Mid-infrared (IR) wavelengths is especially attractive spectral range as it offers a large set of unique and technologically relevant applications \cite{Zhong}. 

Among the diversity of plasmonic materials two-dimensional (2D) structures stand out as especially appealing candidates for plasmonics ~\cite{Low2016,adv2016,Pfnur,C8NR01395K}.
For example, graphene, the most known 2D material, exhibits in many ways unique optoelectronic properties, showing high energy confinement and large tunability \cite{Geim,koppens2011,Fei,chen2012,Fei2012,Grigorenko2012,kim2012,brar2013,StauberReview,Rodrigo2017,yao2018}. 
Intensive research has also been focused on other two-dimensional materials. Among them are transition-metal dichalcogenides and black phosphorus~\cite{Stauber,wang2015,mishra2016,LowBP,Serrano,locplasm}. The former exhibits plasmon resonances in the visible and near ultraviolet ranges, while the latter demonstrates strongly anisotropic optical properties, which makes it suitable for hosting hyperbolic plasmons \cite{Nemilentsau}. On the other hand, emerging 2D materials with magnetic degrees of freedom \cite{Gong,Huang} open up another exciting direction in the field of nanoplasmonics \cite{Armelles}.

In this work, we study plasmon excitations in electron-doped single-layer antimony (SL-Sb), a recently fabricated 2D semiconductor with remarkable environmental stability \cite{Ares2016,Wu2017}, and presumably high carrier mobility \cite{antmob}. Electronic structure of SL-Sb is strongly influenced by the spin-orbit interaction (SOI)~\cite{Zhao2015,Rudenko}, which presumes additional functionalities and control. 
We find that under application of the gate voltage, electron-doped SL-Sb demonstrates unusual low-loss plasmonic excitations in the mid-IR region. The observed excitations are characterized by negative dispersion at small wavevectors, and turn out to be highly tunable by either bias potential or charge doping. The effect mainly originates from the SOI-induced spin-splitting of the conduction band, resulting in the interband resonances, significantly affecting the dielectric response.

\begin{figure}
	\includegraphics[width=\columnwidth]{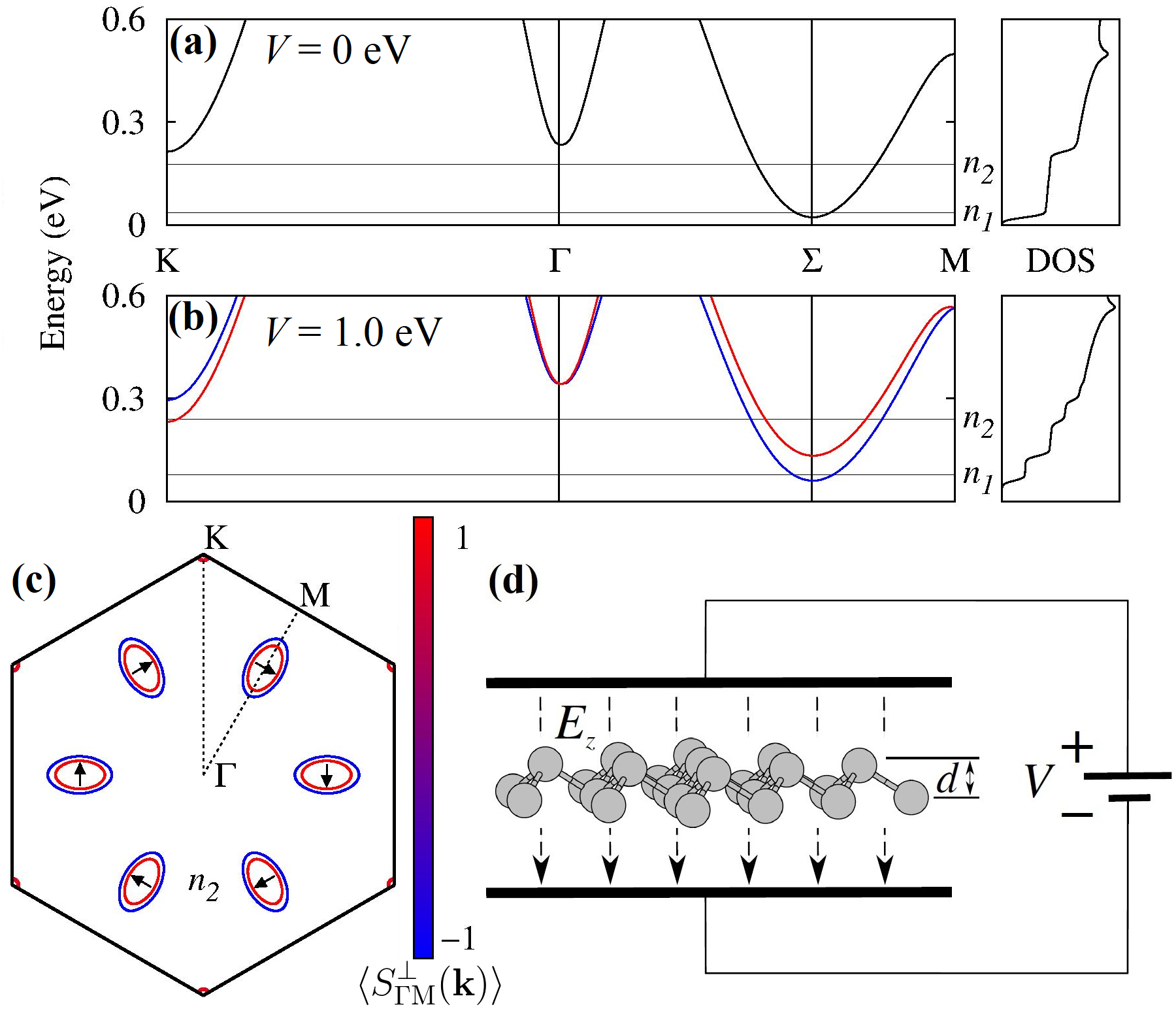}
	\caption{\label{band} Band structure and density of states of SL-Sb calculated in the absence (a), and in the presence (b) of a vertical bias with $V=1$ eV. In each case, black horizontal line marks the Fermi energy corresponding the electron doping with concentrations $n_1=10^{13}$ and $n_2=10^{14}$ cm$^{-2}$. (c) Fermi surface contour for the concentration $n_2$ and bias potential $V = 1$ eV, with colors corresponding to the expectation value (in units of $\hbar/2$) of the spin operator projected on the direction perpendicular to $\Gamma$--M, $S^{\perp}_{\Gamma \mathrm{M}}({\bf k})$. Blue corresponds to the clockwise direction, while red is for the anticlockwise direction. The direction of the total spin per valley is shown by the black arrows. (d) Schematic representation of SL-Sb embedded in electric field controlled by the gate voltage.}
\end{figure}

Antimonene has a hexagonal A7-type crystal structure
(space group $D^3_{3d}$) with the lattice parameter $a$=4.12 \AA~and two sublattices displaced vertically by $b$=1.65 \AA~\cite{Rudenko}.
SL-Sb is predicted to be an indirect gap semiconductor with the gap in the near-IR range~\cite{ant1}. The electronic structure of SL-Sb can be accurately described over a wide energy range using a tight-binding (TB) model proposed in Ref.~\citenum{Rudenko}. The model is defined in the basis of $p$ orbitals and explicitly takes into account SOI. In the presence of a vertical bias the corresponding Hamiltonian has the following form:

\begin{equation}
H=\sum_{ij\sigma \sigma'}t_{ij}^{\sigma \sigma'}c_{i\sigma}^{\dag}c_{j\sigma'}+\frac{V}{d}\sum_{i\sigma}z_{i}c_{i\sigma}^{\dag}c_{i\sigma},
\end{equation}
where $c_{i\sigma}^{\dag}$ ($c_{j\sigma'}$) is the creation (annihilation) operator of electrons with spin $\sigma$ ($\sigma'$) at orbital $i$ ($j$), $z_i$ is the $z$-component of the position operator of the orbital $i$, $t^{\sigma \sigma'}_{ij}$ is the spin-dependent matrix of hopping integrals, $V$ is bias voltage applied to the upper and lower planes of the system, and $d$ is the vertical displacement between the sublattices [Fig.~\ref{band}(d)].

Fig.~\ref{band}(a) shows energy dispersion and density of states (DOS) of the conduction states of SL-Sb. The conduction band minimum corresponds to a low-symmetry $\Sigma$-point ($C_{2v}$ point group), which is located at 0.56~\AA$^{-1}$~from the $\Gamma$-point along the $\Gamma$--M direction of the Brillouin zone. Low energy dispersion at the band edge can be described by two effective masses, $m^{\parallel}_{\Sigma}=0.43m_0$ and $m^{\perp}_{\Sigma}=0.13m_0$, corresponding to the direction along and perpendicular to $\Gamma$--M, respectively. This gives rise to six ellipsoidal valleys formed around the zone center. In the presence of a vertical bias (or perpendicular static electric field), the spin degeneracy is lifted as a result of inversion symmetry breaking \cite{Lugovskoi}. The resulting spin splitting is shown in Fig.~\ref{band}(b), which reaches 0.1 eV at the bias voltage $V=1$ eV. In this situation, the effective masses enhance to $m^{\parallel}_{\Sigma}=0.47m_0$ and $m^{\perp}_{\Sigma}=0.17m_0$ for both bands. The corresponding Fermi energy contours are shown in Fig.~\ref{band}(c), where projections on the opposite spin directions is shown by color. The splitting of electron states in SL-Sb is different from the Rashba splitting typical to narrow gap 2D electron gas, but rather resembles exchange splitting in the ferromagnets \cite{RashbaHamilt}. Indeed, the expectation value of the spin operator projected to the direction perpendicular to $\Gamma$--M, $\langle S^{\perp}_{\Gamma \mathrm{M}}({\bf k}) \rangle = \langle \psi^{\sigma}_i({\bf k})|S^{\perp \sigma \sigma'}_{\Gamma \mathrm{M}}| \psi^{\sigma'}_i({\bf k}) \rangle$, shows that the two states within each valley correspond to the opposite ($\pm \hbar/2$) spin projections [see Fig.~\ref{band}(c)]. In contrast to ferromagnets, time reversal symmetry is preserved in biased SL-Sb, leading to zero net magnetization. 

The combination of a gate-controlled band splitting and finite DOS at the Fermi energy achievable by doping opens up the possibility to tune plasmonic resonances in SL-Sb, which is of interest for practical applications. Here, we restrict ourselves to the case of electron doping only, as it represents the most interesting case. We only note that the properties of hole-doped SL-Sb can be with high accuracy described by the well-studied Rashba model \cite{Rashba}.

To investigate optical properties of SL-Sb, we first calculate frequency-dependent dielectric matrix $\epsilon_{ij}^{(\textbf{q})}(\omega)$. To this end, we use the random phase approximation assuming no dielectric background (free-standing sample):

\begin{equation}
\epsilon_{\sigma \sigma'}^{(\textbf{q})}(\omega)=\delta_{\sigma \sigma'}-\frac{2\pi e^2}{qS}\Pi_{\sigma \sigma'}^{(\textbf{q})}(\omega),
\end{equation}
where $2\pi e^2/qS$ is the long-wavelength approximation of the bare Coulomb interaction density in 2D, and $\Pi_{\sigma \sigma'}^{({\bf q})}(\omega)$ is the polarizability matrix. For the purpose of our study it is sufficient to ignore local field effects related to the orbital degrees of freedom, while the effects of the spin subsystem turn out to be important. Using the definition given above, the spectrum of plasma excitations is determined by the equation $\mathrm{det}[\epsilon_{\sigma \sigma'}^{({\bf q})}(\omega_p)]=i\gamma^{({\bf q})}(\omega_p)$, where $\gamma^{({\bf q})}(\omega_p)$ is the damping factor, and $\omega_p=\omega_p^{({\bf q})}$ is the plasma frequency.
In the spinor basis, the polarizability can be defined as~\cite{graf_electromagnetic_1995}:

\begin{multline}
\begin{split}
\hspace{-0.3cm}&\Pi_{\sigma \sigma'}^{({\bf q})}(\omega)=\sum_{ \substack{ij{\bf k} \\ mn}  }  ( f_{m}^{(\textbf{k})}-f_{n}^{(\textbf{k}')} )  \frac{C_{i\sigma m}^{(\textbf{k})}C^{*(\textbf{k}')}_{i\sigma n}C^{*(\textbf{k})}_{j\sigma' m}C^{(\textbf{k}')}_{j\sigma' n}  }{E_{m}^{(\textbf{k})}-E_{n}^{(\textbf{k}')}+\omega+i\eta} \hspace{-0.45cm} 
\end{split}
\label{Pi}
\end{multline}
where $C_{i\sigma m}^{(\textbf{k})}$ is the contribution of the $i$th orbital $w_{i\sigma}^{{\bf R}}({\bf r})$ with spin $\sigma$ to the Hamiltonian eigenstate 
$\psi_{m}^{{\bf k}}({\bf r})=\sum_{i\sigma {\bf R}} C_{i\sigma m}^{({\bf k})} e^{i{\bf k}\cdot {\bf R}}w_{i\sigma}^{{\bf R}}({\bf r})$
with energy $E_{m}^{({\bf k})}$, ${\bf k}'={\bf k}+{\bf q}$,
$f_m^{({\bf k})}=(\mathrm{exp}[(E_m^{({\bf k})}-\mu)/T]+1)^{-1}$ is Fermi-Dirac occupation factor, $\mu$ is the chemical potential determined by the carrier concentration $n$, and $\eta$ is a broadening term. In our calculations, we used $T=300$ K, $\eta=5$ meV, and two representative values of electron doping, $n=10^{13}$ and $10^{14}$ cm$^{-2}$. Brillouin zone integration has been performed on a grid of $\sim$10$^6$ {\bf k}-points.

\begin{figure}
	\includegraphics[width=\columnwidth]{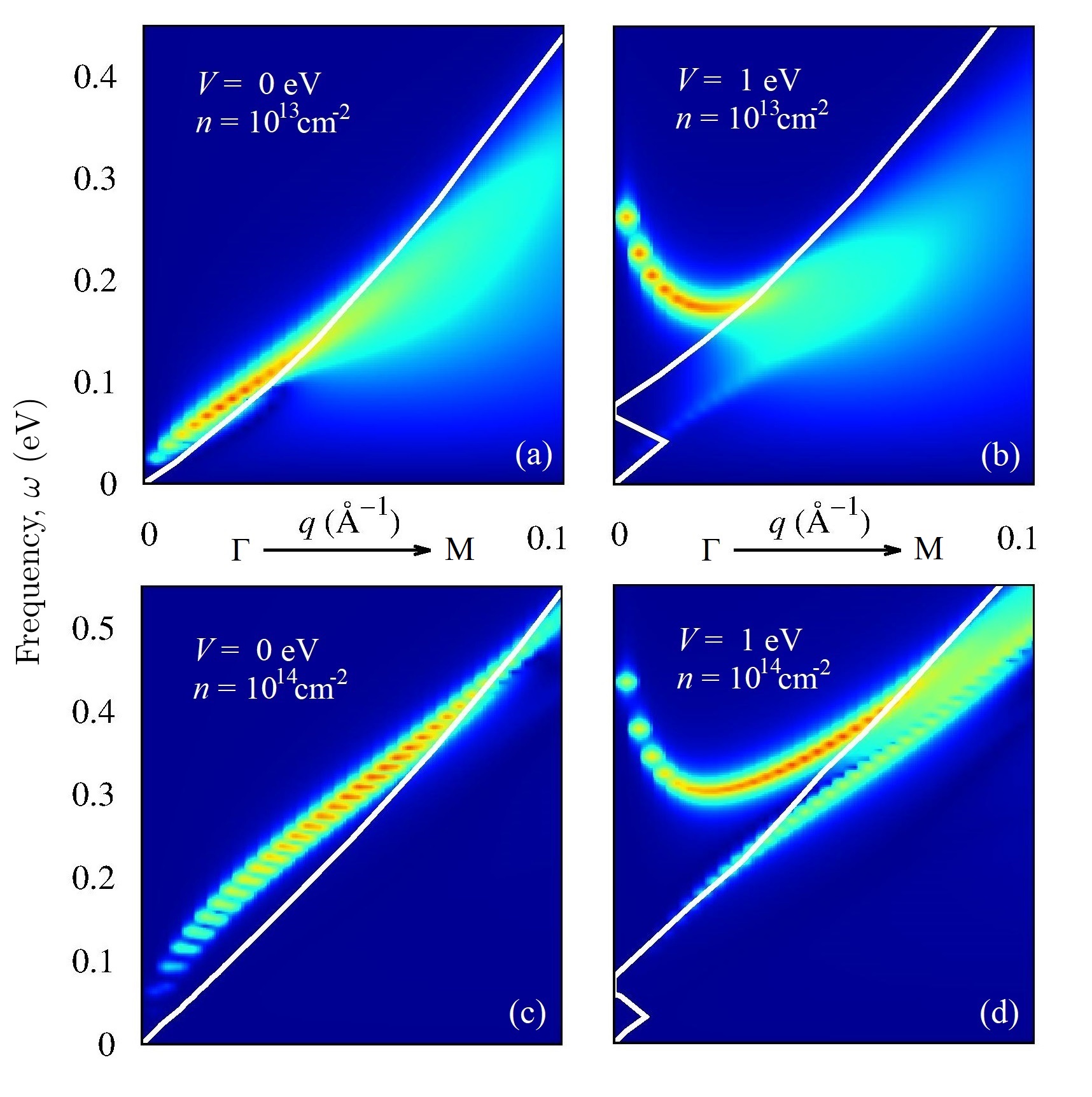}
	\caption{\label{plasm2} Plasmon loss function $L(\textbf{q},\omega)$ in SL-Sb calculated for electron doping concentrations $n=10^{13}$ and 10$^{14}$ cm$^{-2}$ (top and bottom panels), and bias voltages $V=0$ and 1 eV (left and right panels). White line represents boundaries of the particle-hole continuum determined by the poles of the polarization function [Eq.~\ref{Pi}], $\omega_0^{({\bf q})} = \mathrm{max}\{E_n^{({\bf k}+{\bf q})}-E_m^{({\bf k})}\}$.}
    \label{loss}
\end{figure}

To understand the extent to which one can tune the optical properties of SL-Sb, we calculate the plasmon loss function $L(\textbf{q},\omega)=\mathrm{Im}(1/\mathrm{det}[\epsilon^{({\bf q})}_{\sigma \sigma'}(\omega)])$, and study its behavior with respect to the carrier doping and external potential strength. The results are presented in Fig.~\ref{loss}, which also shows boundaries of the particle-hole continuum, $\omega_0^{(\bf q)}$.
In the absence of external potential electron occupy the bottom of a single parabolic band with no interband transition allowed [see Fig.~\ref{band}(a)]. In this situation, optical response is determined by the plasma oscillations of nonrelativistic 2D electron gas, for which one has $\omega_p^2 \approx an|{\bf q}| + bE_Fq^2$, where $a$ and $b$ are constants, and $E_F$ is the Fermi energy \cite{Stern}. The corresponding loss function for $n=10^{13}$ and $10^{14}$ cm$^{-2}$ is shown in Figs.~\ref{loss}(a) and (c), from which one can see a ``classical'' $\sqrt{q}$ plasmon dispersion at low frequencies. At $\omega^{({\bf q})}_p < \omega_0^{({\bf q})}$ the plasmon dispersion enters single-particle excitation continuum and decays into electron-hole pairs.
The energy scale of phonon excitations in SL-Sb lies in the far-IR region \cite{Lugovskoi}, meaning the absence of phonon-plasmon resonances \cite{Low2014} in the relevant spectral range.

\begin{figure}
	\includegraphics[width=\columnwidth]{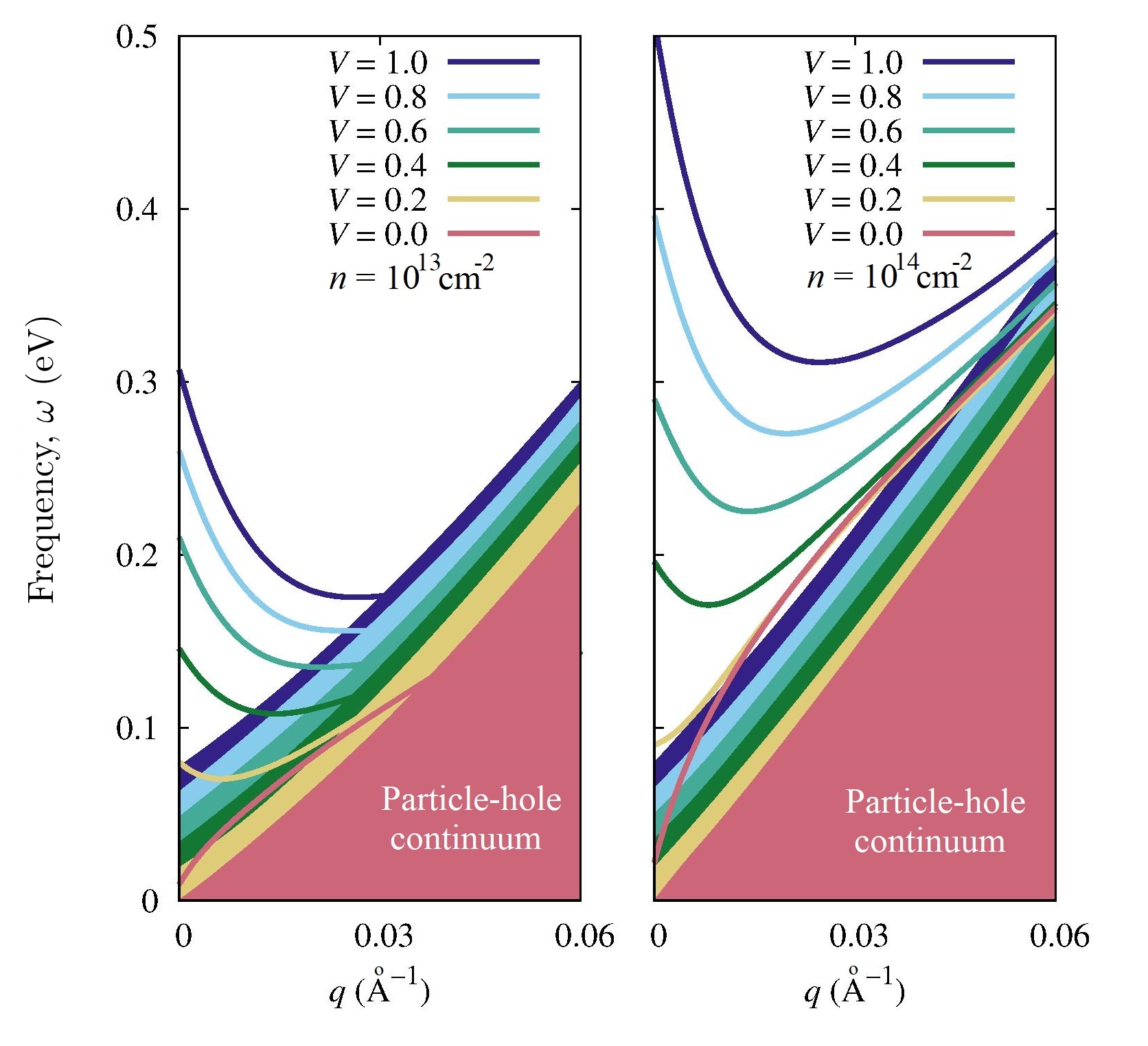}
	\caption{\label{disp} 
Solutions of the equation Re\{$\mathrm{det}[\epsilon^{({\bf q})}_{\sigma \sigma'}(\omega_p)]\}=0$, determining lossless plasmon modes calculated for different bias potentials $V$ (in eV), and for two representative values of electron doping $n$ in SL-Sb. The corresponding region of Landau damping is shown in each case by the same color. For clarity, only the highest (``optical'') plasmon mode is shown at each gate voltage. }
\end{figure}

The plasmon spectrum changes drastically when we introduce external bias potential with magnitude 1 eV, see Figs.~\ref{loss}(b) and (d). In this case, a second (``optical'') plasmon branch appears. The new branch has large spectral weight and lies in the mid-IR region, independently of the electron concentrations considered.
These excitations have a peculiar parabolic-like negative dispersion at small $q$. Their origin is related to the SOI-mediated splitting of the conduction band [Fig.~\ref{band}(b)] allowing for the interband transitions, reminiscent to that in bilayer graphene \cite{Low2014,Gamayun}.
The frequency of the excitations at small $q$ can be effectively tuned by gate voltage, as it is shown in Fig.~\ref{disp}.
Depending on the level of electron doping (10$^{13}$ or 10$^{14}$ cm$^{-2}$), one can smoothly tune $\omega_p(q\rightarrow0)$ from 0 to 0.3 (or 0.5) eV by applying bias voltage in the range up to 1 eV. In all cases relevant excitations lie above the Landau damping region ($\omega_p > \omega_0$), indicating fully coherent plasmon modes. On the contrary, ``classical'' plasmon mode in biased SL-Sb falls inside the particle-hole continuum, and turns out to be essentially damped. As a consequence of the Kramers-Kronig sum rule \cite{KK-sum}, the entire spectral weight at long wavelengths is transferred to the ``optical'' mode. We note that further tunability toward lower frequencies can be achieved by the dielectric substrate (not considered here). 

To gain further insights into the origin of plasma excitations in SL-Sb, we analyze effective dielectric functions $\epsilon^{({\bf q})}_{\mathrm{eff}}(\omega)=\mathrm{det}[\epsilon_{\sigma \sigma'}^{({\bf q})}(\omega)]$, shown in Fig.~\ref{det} for $n=10^{13}$ cm$^{-2}$. Without bias potential [Fig.~\ref{det}(a)] one has typical behavior $\epsilon_{\mathrm{eff}}(\omega) \approx 1-\omega_p^2/(\omega^2+i\omega\gamma)$ at large enough frequencies with $\omega_p^2\sim q$. At $\omega < \omega_p$ there is another solution of the equation $\mathrm{Re}[\epsilon^{({\bf q})}_{\mathrm{eff}}(\omega)]=0$ with $\omega \sim q$. This solution is known as the ``acoustic'' plasmon mode corresponding to out-of-phase charge density oscillations observed in 2D materials with finite thickness, including bilayer graphene \cite{Hwang}, transition metal dichalcogenides \cite{andersen_plasmons_2013}, and phosphorene \cite{BP-plasmons}. Similar to the other systems, this mode is strongly damped as it lies in the particle-hole continuum.
If we introduce bias potential [Fig.~\ref{det}(b)], $\epsilon_{\mathrm{eff}}(\omega)$ exhibits a discontinuity at $\omega=\omega_0(V)$, and has the characteristic shape typical to a conductor with resonant scatterers \cite{Marder},
\begin{equation}
\epsilon_{\mathrm{eff}}(\omega) \approx 1 - \sum_l \frac{\omega_{p,l}^2}{\omega^2-\omega_{0,l}^2+i\omega\gamma_l},
\label{eps_eff2}
\end{equation}
where $l$ denotes different scattering channels, which in our case can be associated with intraband and interband transitions.
Eq.~(\ref{eps_eff2}) allows for the existence of plasmons with negative dispersion if there is $l$ for which $\partial \omega^2_{0,l} / \partial q <0$.
As can be seen from Figs.~\ref{loss}(b) and (d), this condition is fulfilled in biased SL-Sb at $q \lesssim 0.02$ \AA$^{-1}$, which coincides with the region of negative plasmon dispersion. Apart from this prominent solution, there is an overdamped plasmon mode at frequencies close to the interband resonance [Fig.~\ref{det}(b)], while the ``acoustic'' branch turns out to be fully suppressed at large enough $V$. 

\begin{figure}[tbp]
	\includegraphics[width=0.8\columnwidth]{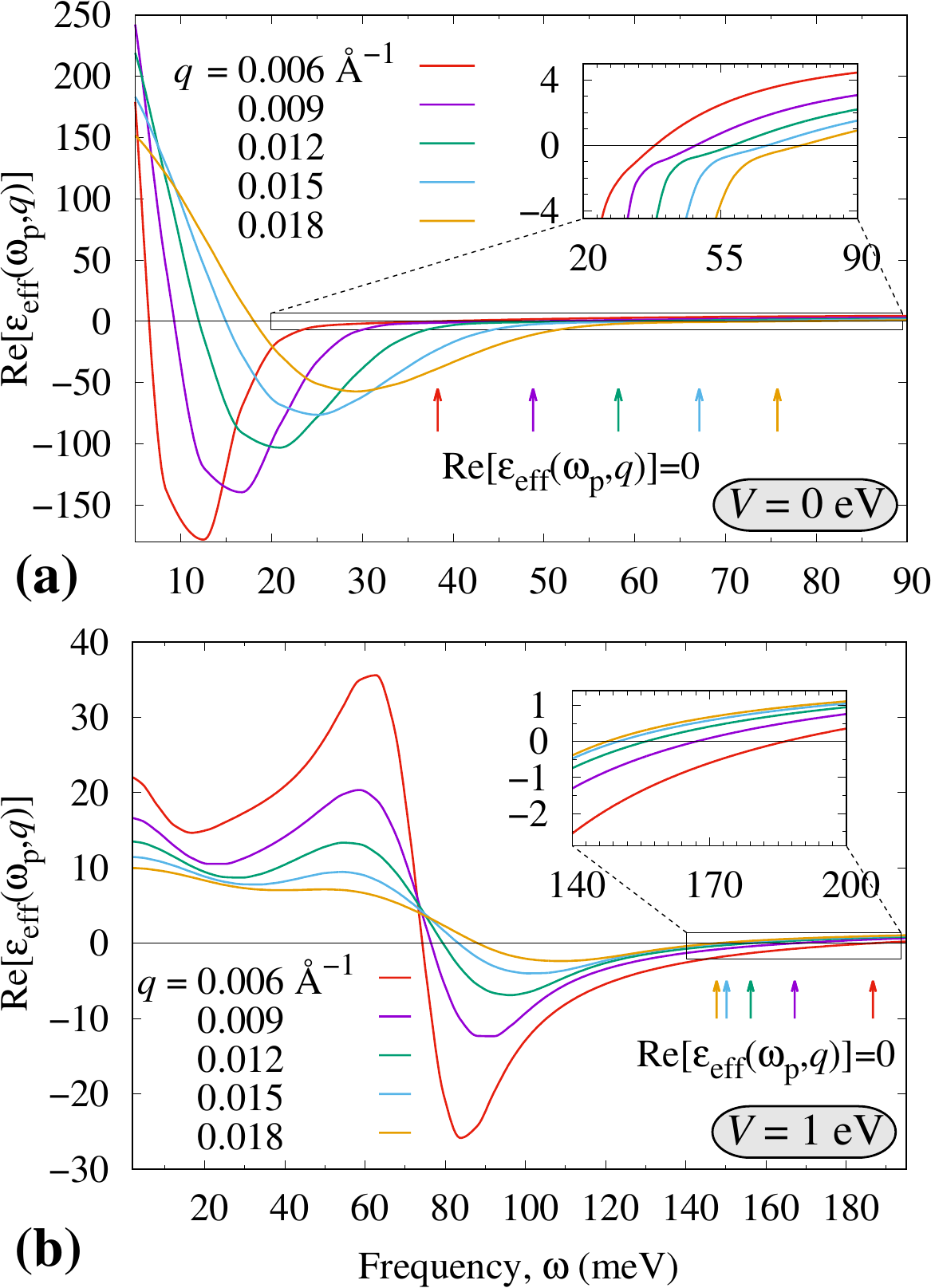}
	\caption{\label{det} Frequency dependence of $\mathrm{Re}[\epsilon_{\mathrm{eff}}(\omega,q)]$ in SL-Sb calculated for a series of {\bf q}-points along the $\Gamma$--M path. (a) and (b) correspond to $V = 0$ and $V = 1$ eV, respectively. In both cases, electron doping of 10$^{13}$ cm$^{-2}$ is assumed. The inset shows a zoom-in of the region where $\mathrm{Re}[\epsilon_{\mathrm{eff}}(\omega,q)]=0$, determining the ``optical'' plasmon modes. Arrows point to the corresponding solutions for the given set of {\bf q}-points.}
    \label{det}
\end{figure}

Plasmon excitations with negative dispersion is uncommon but not unique phenomenon. It was first appeared in the context of bulk Cs crystal \cite{Felde}, but was further observed in other materials \cite{Schuster,Wezel}. A recent study reports similar behavior in electron-doped monolayer MoS$_2$ \cite{MoS2}. Negative dispersion is associated with negative group velocity, indicating negative energy flux. This phenomenon gives rise to an intriguing subfield of nanoplasmonics with a number of exotic optical effects including negative refraction index \cite{Agranovich,Feigenbaum,Compaijen}. 
The fact that the corresponding frequencies in SL-Sb fall in the technologically relevant spectral range make this system prospective for further experimental studies. Plasmons in 2D systems can be accessed by a variety of methods, including electron energy-loss spectroscopy \cite{Geim}, IR optical measurements \cite{Fei}, and scanning probe microscopy \cite{Fei2012}, performed earlier for graphene. High tunability of plasmon excitations in SL-Sb offered by the strong SOI is another appealing aspect to be explored in the context of nanoplasmonic applications.
To experimentally observe the peculiar character of plasmons in SL-Sb, strong electric fields on the order of 0.1--0.5 eV/\AA~may be required. This can be achieved, for example, by the encapsulation of SL-Sb in polar semiconductors \cite{semicond}, or by means of heavy alkali metal doping \cite{ScienceBP}.


To conclude, we theoretically studied optoelectronic properties of SL-Sb at realistic electron concentrations by varying the applied gate voltage.
In addition to the classical 2D plasmon, we find that SOI-induced spin splitting gives rise to a new lossless plasmon branch in the mid-IR region at frequencies highly sensitive to the bias voltage. Remarkably, the new excitations exhibit negative dispersion in a wide range of wavevectors. This behavior is attributed to the strong SOI and inversion symmetry breaking, as well as indicates an important role of the local field effects in the spin channel. Our findings suggest SL-Sb to be an appealing nanoplasmonic material with great gate-tunability, which paves the way for further experimental and theoretical studies in this field.


\begin{acknowledgments}

This work was supported by the Russian Science Foundation, Grant No. 17-72-20041. 
Part of the research was carried out using high performance computing resources at Moscow State University~\cite{lomonosov}.

\end{acknowledgments}

\bibliographystyle{titlenocomma}
\bibliography{apstemplate}


\end{document}